\documentclass[12pt,preprint]{aastex}

\usepackage{rotating}

\begin{document}

\title{Physical Properties and Evolutionary States of EA-type Eclipsing Binaries Observed by LAMOST}

\author{Qian S.-B.\altaffilmark{1,2,3,4}, Zhang J.\altaffilmark{1,2,3}, He J.-J.\altaffilmark{1,2,3},
Zhu L.-Y.\altaffilmark{1,2,3,4}, Zhao E.-G.\altaffilmark{1,2,3,4}, Shi X.-D.\altaffilmark{1,2,3}, Zhou X.\altaffilmark{1,2,3} and Han Z.-T.\altaffilmark{1,2,3,4}}

\altaffiltext{1}{Yunnan
Observatories, Chinese Academy of Sciences (CAS), P.O. Box 110, 650011
Kunming, P. R. China (e-mail: qsb@ynao.ac.cn)}

\altaffiltext{2}{Key laboratory of the structure and evolution of
celestial objects, Chinese Academy of Sciences, P.O. Box 110, 650011
Kunming, P. R. China}

\altaffiltext{3}{Center for Astronomical Mega-Science, Chinese Academy of Sciences, 20A Datun Road, Chaoyang District, Beijing, 100012, P. R. China}

\altaffiltext{4}{University of the Chinese Academy of
Sciences, Yuquan Road 19\#, Sijingshang Block, 100049 Beijing, P. R.
China}

\begin{abstract}

About 3196 EA-type binaries (EAs) were observed by LAMOST by June 16, 2017 and their spectral types were derived. Meanwhile stellar atmospheric parameters of 2020 EAs were determined. In the paper, those EAs are catalogued and their physical properties and evolutionary states are investigated. The period distribution of EAs suggests that the period limit of tidal locking for the close binaries is about 6\,days. It is found that the metallicity of EAs is higher than that of EWs indicating that EAs are generally younger than EWs and they are the progenitors of EWs. The metallicities of long-period EWs ($0.4 < P < 1$ days) are the same as those of EAs with the same periods, while their values of Log (g) are usually smaller than those of EAs. These support the evolutionary process that EAs evolve into long-period EWs through the combination of angular momentum loss (AML) via magnetic braking and case A mass transfer. For short-period EWs, their metallicities are lower than those of EAs, while their gravitational accelerations are higher. These reveal that they may be formed from cool short-period EAs through AML via magnetic braking with little mass transfer. For some EWs with high metallicities, they may be contaminated by material from the evolution of unseen neutron stars and black holes or they have third bodies that may help them to form rapidly through a short timescale of pre-contact evolution. The present investigation suggests that the modern EW populations may be formed through the combination of aforementioned mechanisms.
\end{abstract}


\section{Introduction}

\begin{figure}
\begin{center}
\includegraphics[angle=0,scale=0.6]{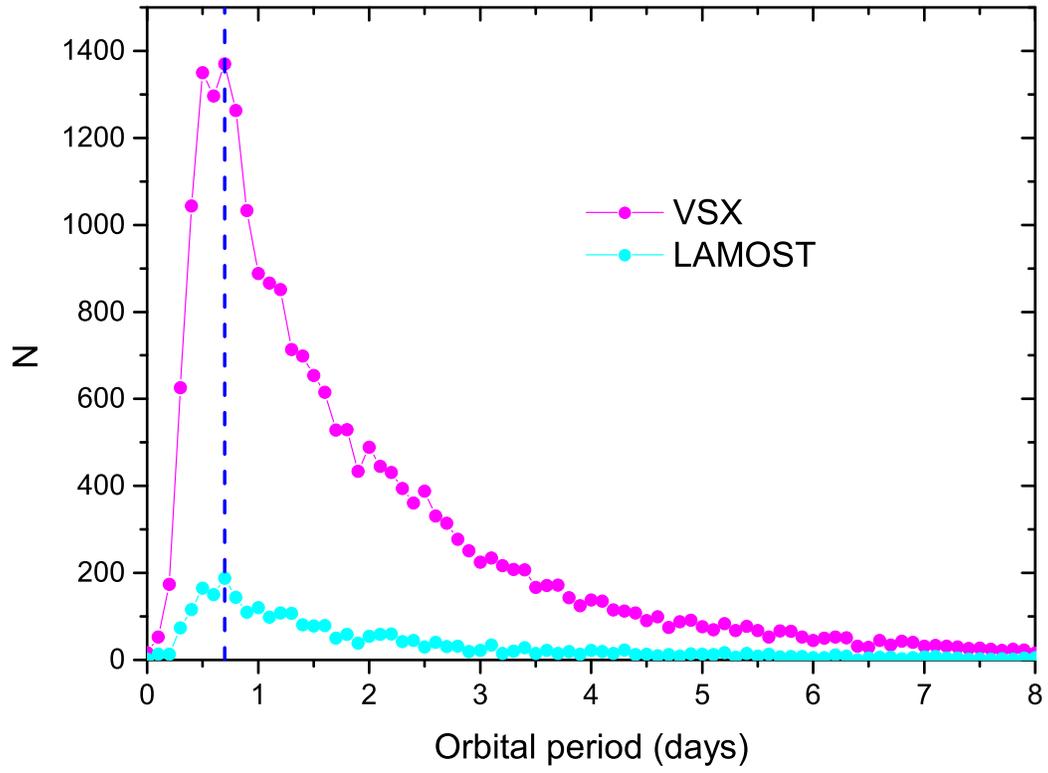}
\caption{Period distribution of EAs observed by LAMOST (solid cyan dots). Also shown as magenta ones are EA systems listed in VSX catalogue. The blue dashed line represents the peak of the two period distributions near 0.7\,days.}
\end{center}
\end{figure}

Eclipsing binaries are very important source in astrophysics because their parameters could be determined reliably based on the photometric light curves and radial velocity curves. In the general catalogue of variable stars (GCVS), according to the shapes of light curves, they are classified in three groups, i.e., EA-, EB- and EW-types (EB$=\beta$-Lyrae type and EW$=$Ursae-Majoris type, Samus et al. 2017). EA-type eclipsing systems (EAs) are close binaries with spherical or slightly
ellipsoidal stellar components. It is possible to specify the moments of the beginning and the end of the eclipses on their light curves. The properties indicate that both components are not in contact with each other. Out of the eclipses, the light remains almost constant or varies insignificantly that is caused by reflection effects and slight ellipsoidality
of binary components, or physical variations. Light amplitudes are also quite different and may reach several magnitudes indicating that there is a large temperature difference between both components (e.g., Samus et al. 2017). The prototype of EA systems is Algol ($\beta$ Per) that is a semi-detached binary. However, many eclipsing binaries with EA-type light curves have turned out to be detached systems. Therefor, EAs listed GCVS are usually detached binary systems and also include classical Algols.

A large number of EAs were discovered by a few big photometric surveys in the world, such as Catalina Sky Survey (CSS, Drake et al. 2009, 2014), the asteroid survey LINEAR (Palaversa et al. 2013), All Sky Automated Survey (ASAS, Pojmanski 1997, Pojmanski et al. 2005) and Northern sky variability survey (NSVS, Wozniak et al. 2004). {\bf Other wide-field surveys which have identified EAs are} Kepler space telescopes (Borucki et al. 2010), the K2 mission (Howell et al. 2014), the HATNet survey (Bakos et al. 2004), SuperWASP (Pollacco et al. 2006), and the KELT survey (Pepper et al. 2012). 20489 EAs were listed in the international variable star index (VSX) that is to bring all of that new information together in a single data repository and provides the tools necessary for the controlled and secure revising of the data (e.g., Watson 2006). However, most of them have not been observed spectroscopically and their spectroscopic properties are unknown because of the lack of big spectroscopic surveys.

To understand the formation and evolution of EW-type contact binaries, some investigators have considered the evolutionary connection between EAs and EWs (e.g., Guinan \& Bradstreet 1988; Bradstreet \& Guinan 1994; Bilir et al. 2005; Gazeas \& Niarchos 2006; Eker et al. 2006, 2008; Yildiz 2014).
The decaying rotation rate for stars with spectral types later than F was detected by Skumanich (1972) that is explained as
angular momentum loss (AML) through magnetically driven stellar winds, i.e., the magnetic braking (e.g., Kraft 1967). When the late-type stars as a component in a short-period close binary, the strong tidal interaction drives the orbital angular momentum to supply the AML. Therefore, the
AML from the cool components in the tidally locked binaries causes the orbit to shrink and then spins up the components by synchronization. In this way, the initially short-period detached EAs are evolving into EWs (e.g., Huang 1966; Guinan \& Bradstreet 1988; Maceroni \& van¡¯t Veer 1991; Demircan et al. 2006).

During the aforementioned evolutionary process, both the orbital shrinkage due to AML and the component expands via nuclear reaction in their cores could cause the more massive component to fill the critical Roche lobe and transfer mass to its companion. Therefore, the evolution of EAs and the formation of EWs should be a combination of AML and mass transfer. For short-period cool EWs such as BI Vul and CSTAR038663, they may been formed by AML with little mass transfer (e.g., Qian et al. 2013a, 2014). By analyzing stellar atmospheric parameters of 5363 EWs, Qian et al. (2017a) found that the physical properties and evolutionary states of EWs are mainly depending on their orbital periods. Field EWs were divided into several groups by Bilir et al. (2005) based on their orbital period. It is possible that the formation scenarios for different groups of EW are quite different. The kinematical ages of shorter-period
less-massive systems are longer than their long-period cousins. This is in agreement with the conclusion derived by Qian et al. (2017a) who found that
short-period EWs usually have lower metallicities than those of their long-period cousins indicating that they may be older and have a longer timescale of pre-contact evolution. Their formation and evolution are mainly driven by angular momentum loss via magnetic braking. Moreover, EWs have the lowest angular momentum and the shortest orbital period among main-sequence binaries, third bodies may play an important role for their formation and evolution by removing angular momentum during early dynamical interaction and/or later evolution (e.g., Pribulla
\& Rucinski 2006; Qian et al. 2014; Zhu et al. 2013a). These results may indicate that the formation of EWs are complex and debates on the formation mechanisms for EWs continue (e.g., Eker et al. 2008). A detailed investigation on LAMOST spectroscopic data of EAs and a comparison of their physical properties with those of EWs are required.

Among the 25364 EAs listed in VSX by July 16, 2017, about 3196 systems (about 12.6\%) were observed by LAMOST survey from October 24, 2011 to June 16, 2017. The LAMOST is a large sky area multi-object fiber spectroscopic telescope (also called as Guo Shou Jing telescope). It is a special telescope that could obtain the spectra of about 4000 objects simultaneously in an exposure (Wang et al. 1996; Cui et al. 2012). The spectral resolution is about 1800. Huge amounts of spectroscopic data were obtained (e.g., Zhao et al. 2012, Luo et al. 2012, 2015) including close binaries and pulsating stars (e.g., Qian et al. 2017a, b). We catalogued LAMOST observations for 7938 EW-type eclipsing binaries (EWs) and investigate their statistical properties and evolutionary states (Qian et al. 2017a). In the paper, we catalogue the LAMOST data of the 3196 EAs and present their spectroscopic properties. The period distribution for the observed EAs by LAMOST is shown in Fig. 1 as solid dots where 66 EAs without orbital periods are not displayed. Also shown in the figure is the period distribution of all EAs in VSX (open circles). The spectral types and the stellar atmospheric parameters for those EAs observed by LAMOST survey are catalogued firstly. Then, the distributions of those atmospheric parameters and some statistical correlations are given. Finally, by comparing the distributions and relations of the observed EAs with those of EWs, the physical properties and evolutionary states of EA binaries are discussed.

\section{Catalogue of EAs observed by LAMOST}

\begin{figure}
\begin{center}
\includegraphics[angle=0,scale=0.6]{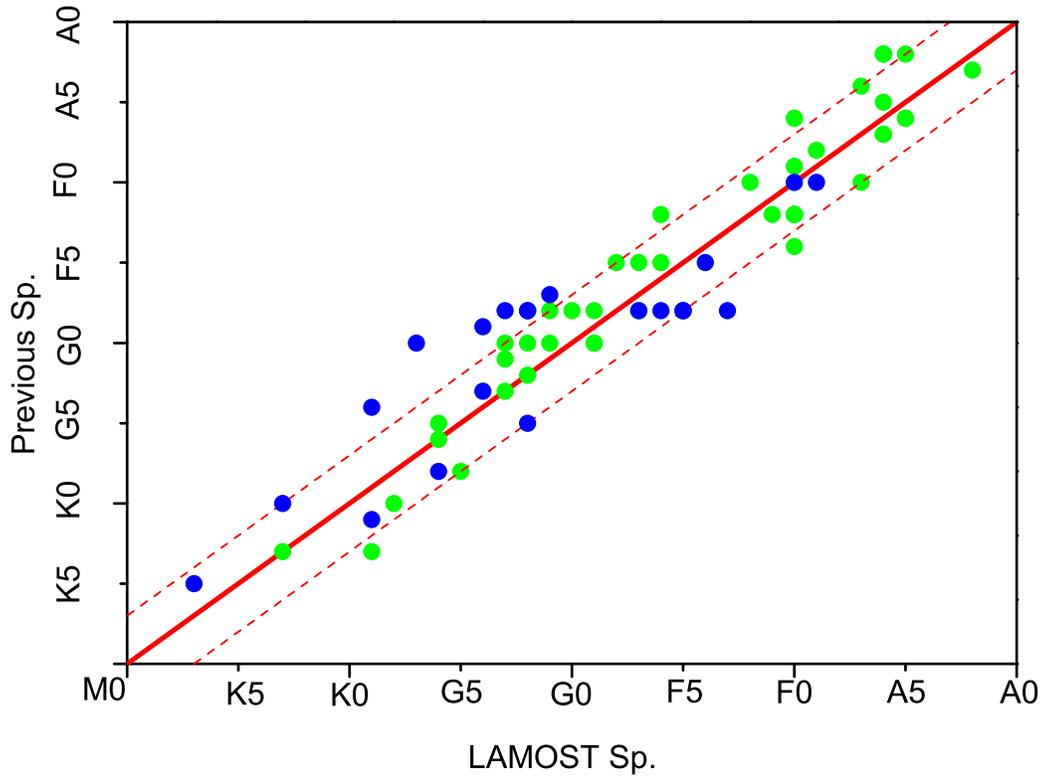}
\caption{Comparison of spectral types for 51 EA-type eclipsing binaries (green dots) that were observed by LAMOST and by previous investigators. Also shown as blue dots are EW-type contact binary stars. It is shown that most of them are in agreement with each other within three subclasses (red dashed lines).}
\end{center}
\end{figure}

\begin{figure}
\begin{center}
\includegraphics[angle=0,scale=0.6]{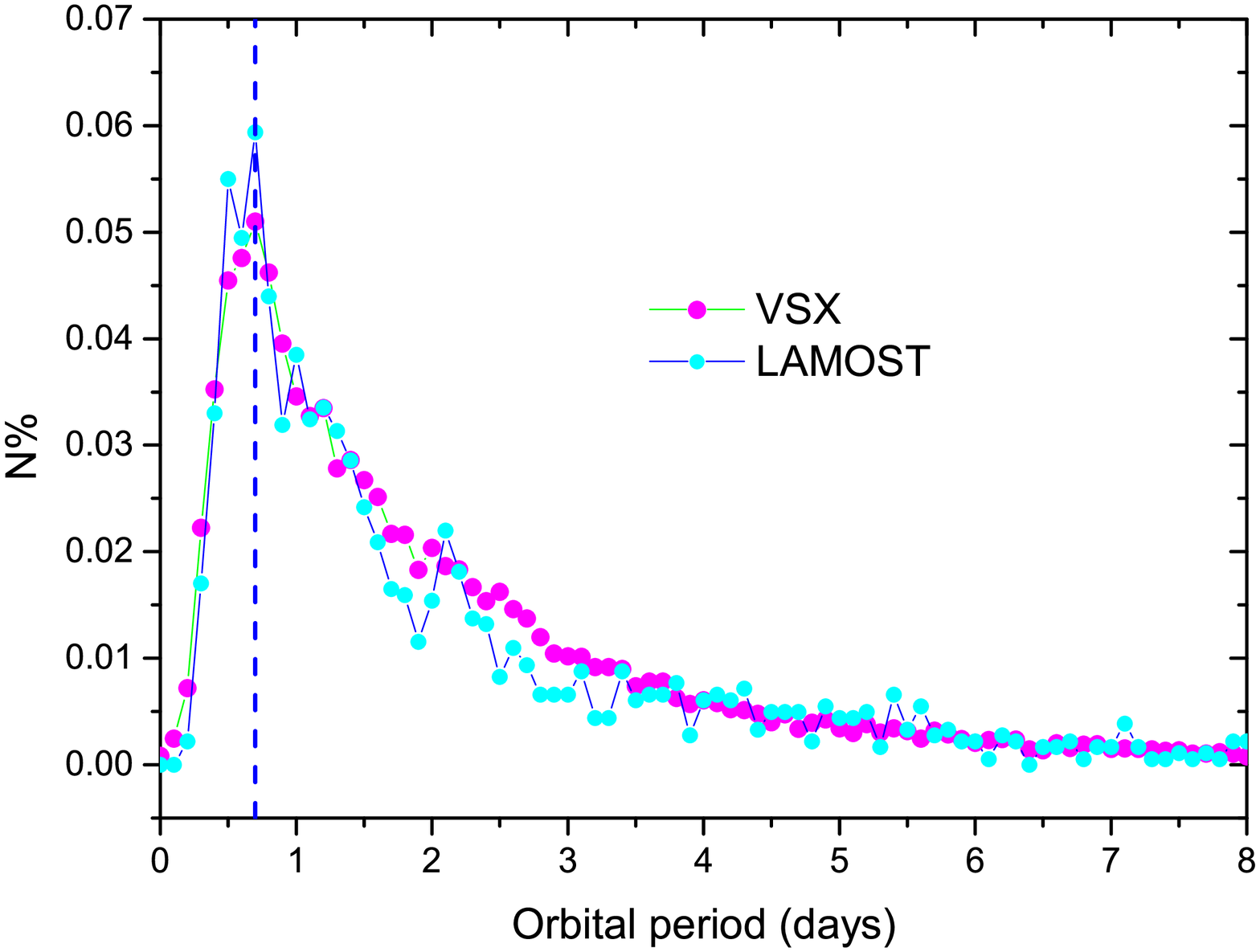}
\caption{Comparison of relative distribution of orbital period for EAs. Symbols are the same as those shown in Fig. 1. The orbital periods are taken from VSX.}
\end{center}
\end{figure}

About 3196 EAs shown in the VSX catalogue were observed by LAMOST in the time interval between October 24, 2011 and June 16, 2017. When the spectra have higher signal to noise, the stellar atmospheric parameters, i.e, the effective temperature $T$, the gravitational acceleration Log g, the metallicity [Fe/H] and the radial velocity $V_{r}$ were determined.
The SNR cutoff is 6 in the g-band from dark nights (eight nights before and after the new moon) observations or 20 from bright nights (neither close to new nor full moon), for those greater SNRs the parameters (e.g., T, Log g and [Fe/H]) were determined. Among the EAs listed in VSX, the stellar atmospheric parameters of 2020 systems were automatically derived by the LAMOST stellar parameter pipeline (e.g., Wu et al. 2011b, 2014; Luo et al. 2015). As in the cases of EWs (Qian et al. 2017), those stellar atmospheric parameters were determined based on the Universite de Lyon spectroscopic analysis software (ULySS) (e.g., Koleva et al. 2009, Wu et al. 2011a).
However, binary systems contain (at least) two component stars (the primary and the secondary) with two different temperatures (as well as Log g and [Fe/H)). The pipeline measures their spectra as a single stars, so only one set of parameters were obtained for the primary components. Does the measured single temperature can represent the primary temperature generally, and what is the typical difference between them? Liu et al. (2017) did an experiment to answer that question. They combined two single stars¡¯ spectra together, and measure the combined spectra as they are single stars with the same pipeline. The binary spectra are found very subtle difference with single stars. The deviations on temperatures between the primary stars and the single measured values are mostly less than 200 K, but for metal rich stars the deviations will reach to 200--500\,K. As for the gravitational acceleration Log g and the metallicity [Fe/H], their deviations are mostly smaller than 0.2 dex. The systematic bias depends on binary parameters strongly, such as metallicity, mass ratio, and temperature difference, etc. For binaries with large temperature differences, small mass ratios, or low metallicities, the systematic biases are usually small. When the temperature differences are around 250\,K, the systematic biases reach their maxima which are coarsely $\sim200$\,K for T, $\sim0.15$ for Log g and [Fe/H]. However, when the temperature differences are larger than 1000\,K, the systematic biases will be less than half of the above values. The mass ratios mainly affect the temperature biases, making T deviations from 200\,K to 50\,K when the mass ratio changes from 1 to 0.25. As for the influence of [Fe/H], the systematic biases will decrease from
$\sim250$\,K for T, $\sim0.25$ for Log g and $\sim0.15$ for [Fe/H] to less than one third of those values when [Fe/H] declines from 0.5 to -2.0.

During the LAMOST survey, 62 EAs were observed four times or more and their atmospheric parameters were averaged. Both the mean values and the corresponding standard errors are listed in Table 1.
The star name, the orbital period and the observational times are listed in the columns 1-3 of the table.
The averaged parameters and their standard errors are listed in the rest columns. As shown in the table, apart from six EAs, the standard errors of the effective temperature for the rest targets are lower than 100\,K. The standard errors of the gravitational acceleration Log g for most EAs are lower than 0.1\,dex. Apart from four EAs, the standard errors of the metallicity for the rest EAs are lower than 0.1\,dex.

Among the 3196 EAs, some of them have been observed by previous investigators. To check the reliability of the spectral data observed by LAMOST, it is good to compare the spectral types observed by both LAMOST and previous authors. We have collected spectral types of 51 EAs and are shown in Fig. 2. Most of the spectral types are from the GCVS (e.g., Samus et al. 2017). The spectral types of CV Boo, CO And, KU Aur, DU Leo and EH Peg were determined by Popper (1996). For the rest targets, UV Leo, VZ CVn, HS Aur, FL Lyr, RX Gem and UX Boo, the spectral types were published by
Popper (1997, 1988), Popper et al. (1986) and Samus' et al. (2017), respectively. For comparison, the spectral types of {\bf 21 EWs} are displayed in the figure as green dots. More than 100 EWs were observed spectroscopically by using the 1.88-m telescope at the David Dunlap Observatory (DDO) of the University of Toronto. Among those EWs, eighteen were also observed by the LAMOST. They are AQ Psc, UV Lyn (Lu \& Rucinski 1999), UZ Leo, AH Aur, XZ Leo (Rucinski \& Lu 1999), AO Cam, UX Eri, CN And (Rucinski et al. 2000), FI Boo (Lu et al. 2001), ET Leo (Rucinski et al. 2002), QW Gem (Rucinski et al. 2003), TX Cnc (Pribulla et al. 2006), XY Leo, AM Leo, CC Com (Pribulla et al. 2007), PY Vir (Rucinski et al. 2008), AU Ser and QX And (Pribulla et al. 2009a).
Also shown in the figure are four other EWs AL Oph (Bond \& Tifft 1974), AW UMa (Pribulla \& Rucinski 2008), DN Boo (Senavci et al., 2008) and TY UMa (Li et al., 2015). As displayed in Fig. 2, {\bf 16 EAs agree within 1 subclass, while 44 ones agree within 3 subclasses. The rest 7 EAs are not consistent within 3 subclasses. Among the 21 EWs, 14 of them agree within 3 subclasses. 7 EWs show} slightly large differences that may be caused by the influences of rapidly rotating and highly deformed component stars in a common envelope.

\begin{table}[!h]
\scriptsize
  \caption{Mean atmospheric parameters and their standard errors for 62 EAs observed four times or more.}
  \begin{center}
\begin{tabular}{llccccccc}\hline
 Star Name              &     P (days)  &  Times  &   $\overline{T}$(K) &   Errors &   $\overline{Log (g)}$   &  Errors   &   $\overline{[Fe/H]}$ & Errors  \\\hline
    NU Leo                         & 1.297914 & 7     & 6690  & 71.18 & 4.15  & 0.033 & -0.27 & 0.049 \\
    NV UMa                         & 3.02405 & 7     & 6237.17 & 40.35 & 4.173 & 0.026 & -0.469 & 0.025 \\
    KID 06847018                   & 16.66213 & 6     & 6234.6 & 44.66 & 4.351 & 0.039 & -0.1  & 0.074 \\
    CSS\_J071343.4+481534           & 2.35406 & 6     & 6413.33 & 130.95 & 4.082 & 0.078 & -0.207 & 0.044 \\
    CSS\_J072637.0+271013           & 0.85506 & 6     & 5736.67 & 70.33 & 3.983 & 0.091 & 0.06  & 0.07 \\
    WX And                         & 3.001134 & 5     & 7250  & 101.49 & 4.064 & 0.015 & 0.092 & 0.086 \\
    ROTSE1 J193141.03+464844.1     & 1.382614 & 5     & 7266  & 76.35 & 3.87  & 0.062 & 0.18  & 0.021 \\
    CSS\_J010411.8+074715           & 0.92412 & 5     & 6434.44 & 85.82 & 4.254 & 0.076 & -0.701 & 0.093 \\
    ASAS J184844+4342.4            & 0.580729 & 5     & 6650.85 & 20.34 & 4.149 & 0.011 & 0.117 & 0.013 \\
    V0523 And                      & 0.52854 & 5     & 5028.53 & 48.89 & 4.435 & 0.058 & 0.232 & 0.037 \\
    CSS\_J073855.7+352208           & 0.905827 & 5     & 5594  & 35.07 & 4.348 & 0.034 & -0.406 & 0.018 \\
    CSS\_J012312.6-023926           & 0.655281 & 5     & 5296  & 250.76 & 4.4   & 0.158 & -0.252 & 0.169 \\
    KID 07943535                   & 4.719294 & 5     & 6657.07 & 33.9  & 4.163 & 0.02  & -0.245 & 0.026 \\
    KID 08019043                   & 1.985608 & 5     & 6468  & 17.89 & 4.252 & 0.016 & -0.246 & 0.021 \\
    KID 09833618                   & 1.408565 & 5     & 6527.19 & 50.97 & 3.967 & 0.054 & 0.128 & 0.047 \\
    KID 10978178                   & 2.27447 & 5     & 7182  & 16.43 & 3.988 & 0.046 & -0.072 & 0.038 \\
    KID 07601633                   & 15.66325 & 5     & 6061.27 & 47.9  & 3.98  & 0.055 & 0.068 & 0.028 \\
    KID 11189127                   & 2.671973 & 5     & 5897.36 & 102.67 & 4.454 & 0.12  & 0.319 & 0.059 \\
    KIC 6850665                    & 14.71606 & 5     & 4939.02 & 32.02 & 2.944 & 0.048 & -0.177 & 0.022 \\
    KIC 8747222                    & 1.667374 & 4     & 4670  & 14.14 & 2.862 & 0.029 & -0.178 & 0.03 \\
    V0874 Mon                      & 2.448816 & 4     & 6752.5 & 12.58 & 4.145 & 0.013 & -0.125 & 0.019 \\
    NU UMa                         & 5.50762 & 4     & 5605.05 & 29.46 & 4.641 & 0.03  & 0.288 & 0.015 \\
    CSS\_J050254.0+210237           & 5.82794 & 4     & 6430  & 153.41 & 3.953 & 0.092 & -0.547 & 0.151 \\
    CSS\_J050313.4+190251           & 2.093145 & 4     & 6727.5 & 80.57 & 3.78  & 0.037 & 0.295 & 0.044 \\
    CSS\_J053548.5+144618           & 1.43555 & 4     & 6592.5 & 63.97 & 4.045 & 0.04  & -0.227 & 0.095 \\
    KID 11553706                   & 3.719755 & 4     & 5987.41 & 26.29 & 4.29  & 0.045 & -0.198 & 0.03 \\
    CSS\_J235856.7+371823           & 1.354679 & 4     & 6515  & 34.16 & 4.188 & 0.03  & -0.145 & 0.024 \\
    KID 09818381                   & 6.045496 & 4     & 5942.5 & 34.03 & 4.31  & 0.105 & 0.308 & 0.059 \\
    KID 08669092                   & 1.000804 & 4     & 6212.26 & 20.66 & 4.127 & 0.034 & -0.11 & 0.012 \\
    CSS\_J064736.2+423643           & 1.79118 & 4     & 6235  & 97.47 & 4.02  & 0.055 & -0.055 & 0.082 \\
    ASAS J071149+1950.2            & 1.7695 & 4     & 5517.5 & 47.17 & 4.37  & 0.159 & 0.28  & 0.023 \\
    CSS\_J013541.9+383916           & 1.27451 & 4     & 5012.5 & 90.69 & 4.433 & 0.097 & 0.018 & 0.112 \\
    DV Gem                         & 4.40042 & 4     & 6152.5 & 231.28 & 3.795 & 0.217 & 0.035 & 0.068 \\
    GSC 01394-01889                & 1.470827 & 4     & 6125  & 12.91 & 4.095 & 0.047 & -0.408 & 0.072 \\
    CSS\_J070728.8+410115           & 1.53604 & 4     & 4332.5 & 35.94 & 4.488 & 0.107 & -0.198 & 0.05 \\
    KID 11959569                   &               & 4     & 5315  & 36.97 & 3.88  & 0.05  & 0.352 & 0.022 \\
    KID 05534702                   & 1.025473 & 4     & 6835  & 45.09 & 4.078 & 0.044 & -0.143 & 0.057 \\
    KID 06781535                   & 9.122071 & 4     & 5808.48 & 61.72 & 4.241 & 0.094 & -0.084 & 0.044 \\
    KID 08488876                   & 5.801887 & 4     & 6791.27 & 20.13 & 4.04  & 0.032 & 0.071 & 0.013 \\
    KID 08429450                   & 2.704917 & 4     & 7162.5 & 55.6  & 3.955 & 0.053 & 0.125 & 0.039 \\
    KID 03120320                   & 10.2656 & 4     & 5842.59 & 21.2  & 4.146 & 0.014 & 0.138 & 0.018 \\
    KID 05786154                   & 97.92078 & 4     & 4620.57 & 15.22 & 2.607 & 0.063 & -0.127 & 0.067 \\
    KID 07097571                   & 2.213856 & 4     & 6308.16 & 14.39 & 3.964 & 0.022 & 0.257 & 0.008 \\
    NSVS 2272724                   & 2.03898 & 4     & 6190  & 38.3  & 4.343 & 0.039 & -0.017 & 0.017 \\
    NSVS 7685021                   & 2.281979 & 4     & 5777.5 & 72.28 & 4.21  & 0.127 & -0.007 & 0.049 \\
    NSVS 4854614                   & 1.26034 & 4     & 7357.5 & 72.74 & 4.148 & 0.017 & -0.31 & 0.024 \\
    NSVS 12838110                  & 7.658401 & 4     & 5440  & 58.88 & 3.682 & 0.079 & -0.027 & 0.051 \\
    ASAS J190934+4305.9            & 0.57253 & 4     & 7060  & 24.49 & 4.14  & 0.024 & -0.375 & 0.033 \\
    ASAS J120313+0354.0            & 2.10815 & 4     & 5952.5 & 40.31 & 3.995 & 0.084 & 0.035 & 0.013 \\
    ASAS J184321+4734.7            & 0.734562 & 4     & 7447.5 & 34.03 & 4.2   & 0.014 & -0.103 & 0.033 \\
    T-UMa0-01822                   & 9.5514 & 4     & 6139.92 & 28.21 & 4.283 & 0.022 & -0.239 & 0.015 \\
    CSS\_J092612.7+220754           & 0.398744 & 4     & 4987.5 & 104.36 & 4.395 & 0.167 & -0.478 & 0.075 \\
    GSC 01826-00950                & 3.10479 & 4     & 5950  & 96.61 & 4.23  & 0.182 & -0.033 & 0.044 \\
    T-Lyr1-05114                   & 0.708551 & 4     & 6157.5 & 80.16 & 4.065 & 0.04  & 0.263 & 0.034 \\
    TSVSC1 TN-N030301120-1686-82-2 & 1.559231 & 4     & 6812.5 & 22.17 & 4.055 & 0.013 & -0.163 & 0.013 \\
    CSS\_J085306.9+202732           & 3.24434 & 4     & 4272.5 & 27.54 & 4.648 & 0.07  & -0.21 & 0.134 \\
    CSS\_J085704.9+414618           & 1.3419 & 4     & 6332.5 & 41.13 & 4.13  & 0.027 & -0.55 & 0.039 \\
    V0474 Cyg                      & 23.65881 & 4     & 5977.5 & 55.6  & 4.148 & 0.067 & 0.148 & 0.026 \\
    MW Com                         & 2.1677 & 4     & 6505.73 & 19.66 & 4.091 & 0.091 & -0.239 & 0.068 \\
    KIC 7431665                    & 81.4  & 4     & 4691.62 & 11.14 & 2.803 & 0.039 & -0.065 & 0.013 \\
    NSVS 9977672                   & 0.44767 & 4     & 6282.5 & 80.57 & 4.185 & 0.064 & -0.167 & 0.058 \\
    KID 05596440                   & 10.47486 & 4     & 6511.82 & 9.47  & 4.17  & 0.054 & -0.098 & 0.019 \\
\hline
\end{tabular}
\end{center}
\end{table}

We catalogue 2956 groups of observations for the 2020 EAs in the order of their coordinates. In the catalogue, the binary names, their right ascensions (RA) and declinations (DEC), types of light variation and orbital periods are from the VSX catalogue. Column 6 includes the distances (in arcsec) between the two positions determined by the coordinates given in VSX and by LAMOST. They are used to identify those EAs from the LAMOST samples based on the criterion Dist$<2$\,arcsecs. Those shown in the 7th and 8th columns are the observing dates and the spectral types. The rest eight columns list the stellar atmospheric parameters, $T$, Log g, [Fe/H] and $V_{r}$ for the 2020 EAs and their corresponding errors $E_1$, $E_2$, $E_3$ and $E_4$ respectively. As we have done for EWs (Qian et al. 2017a), some EAs were observed twice or more times on different dates and we show all of the parameters. The first 50 observations are shown in Table 2, while the whole catalogue for EAs is available from the electronic version.

Fig. 3 shows the relative distribution of the orbital period (the percentage of the number to the whole sample) for the 2020 EAs where the orbital periods are taken from VSX. The relative period distribution of all EAs listed in VSX is also shown in the same figure. As we can see that both of the two distributions are nearly overlapping. In order to examine mathematically whether the periods of EAs shorter than eight days from VSX and observed by LAMOST are drawn from the same distribution, we performed three independent statistic tests which are Kolmogorov-Smirnov test, the Kruskal-Wallis H-Test, and the nonparametric Rank-Sum Test. The results show that their probabilities are 0.17, 1.00 and 0.31 which are all greater than the 0.05 significance level, indicating the VSX and LAMOST samples have the same mean of distribution. These results indicate that there are no selected effective for those 2020 EAs and they could be used to represent the physical properties of the whole EAs in the total VSX catalogue. The peak of the period distribution is near 0.7\,days (the dashed lines in Figs. 1 and 2). Just like the cases of EWs, for some EA spectra, their signals to noise are not high enough to determine the stellar atmospheric parameters and only their spectral types were given. The spectral types of those EAs are also catalogued and those listed in Table 3 are the first 50 spectral types in the catalogue. The catalogue shows 2160 spectral types and the whole catalogue is available from the electronic version. The descriptions of the columns are the same as those in Table 2.

\begin{sidewaystable}[h]
\tiny
\begin{center}
\caption{The catalogue of the stellar atmospheric parameters for EAs observed by LAMOST (the first 50 observations).}\label{XXXX}
\begin{tabular}{llllllllllllllll}\hline\hline
Name & R.A. (deg) & Dec. (deg) & Type &  P (days) & Dist & Date & Sp. & T (K)& $E_1$ & Log(g) & $E_2$
&[Fe/H] & $E_3$ & RV & $E_4$\\\hline
     NSVS 3635231                   & 0.14775 & 43.01481 &  EA                               & 5.297 & 0.008 & 2014/12/18 &          F6  & 6240  & 40    & 3.97  & 0.04  & -0.03 & 0.03  & 31    & 3 \\
     NSVS 3681222                   & 0.30953 & 50.11794 &  EA                               & 2.968107 & 0.629 & 2013/11/22 &          F5  & 6420  & 10    & 4.12  & 0.01  & -0.03 & 0.01  & 1     & 1 \\
     KELT-1                         & 0.36217 & 39.38383 &  EA                               & 1.217494 & 0.111 & 2013/11/14 &          F5  & 6590  & 10    & 4.12  & 0     & 0.12  & 0     & -16   & 1 \\
     SERIV 29                       & 0.45358 & 31.96164 &  EA                               & 1.884216 & 0.23  & 2012/11/25 &         A7V  & 7460  & 80    & 4.06  & 0.11  & 0.06  & 0.07  & -34   & 7 \\
     CSS\_J000153.9+323530           & 0.47458 & 32.59178 &  EA                               & 1.63844 & 0.084 & 2011/12/12 &          F0  & 6700  & 270   & 4.31  & 0.39  & -0.37 & 0.25  & -62   & 20 \\
     2MASS J00021000+4748036        & 0.54167 & 47.801 &  EA                               & 9.57487 & 0.042 & 2013/10/30 &          F7  & 6080  & 10    & 3.89  & 0     & -0.03 & 0     & -8    & 1 \\
     2MASS J00023306+3315173        & 0.63775 & 33.25481 &  EA                               & 2.37554 & 0.07  & 2012/11/25 &          F5  & 6330  & 20    & 4.01  & 0.02  & 0.06  & 0.02  & -13   & 2 \\
     2MASS J00023306+3315173        & 0.63775 & 33.25481 &  EA                               & 2.37554 & 0.264 & 2015/9/25 &          F7  & 6310  & 10    & 4.07  & 0     & 0.07  & 0     & -32   & 1 \\
     HIP 247                        & 0.76543 & 8.26129 &  EA                               & 2.2604 & 0.022 & 2016/12/16 &          F5  & 6541.23 & 1.88  & 4.185 & 0.001 & -0.109 & 0.002 & -75.65 & 0.41 \\
     NSVS 3684054                   & 1.05488 & 49.70422 &  EA                               & 1.27228 & 0.053 & 2013/11/22 &          G1  & 5960  & 50    & 3.94  & 0.07  & 0.19  & 0.05  & -75   & 5 \\
     AM And                         & 1.28137 & 48.4515 &  EA                               & 8.85051 & 0.513 & 2013/11/22 &        A3IV  & 8230  & 150   & 2.7   & 0.21  & 0.03  & 0.14  & -17   & 13 \\
     NSVS 9100938                   & 1.46756 & 15.38195 &  EA                               & 8.18417 & 0.568 & 2012/9/28 &          F6  & 6190  & 170   & 4.02  & 0.23  & -0.4  & 0.15  & 0     & 12 \\
     NSVS 6318847                   & 1.62138 & 35.87844 &  EA                               & 0.897843 & 0.109 & 2015/9/25 &          F7  & 6020  & 50    & 4.1   & 0.06  & -0.28 & 0.04  & -85   & 4 \\
     NSVS 6318847                   & 1.62138 & 35.87844 &  EA                               & 0.897843 & 0.077 & 2012/11/30 &          G0  & 6020  & 30    & 4.23  & 0.03  & -0.13 & 0.02  & -22   & 3 \\
     2MASS J00074408+4028354        & 1.93367 & 40.4765 &  EA                               & 2.84981 & 0.097 & 2013/11/14 &          F7  & 6310  & 10    & 4.06  & 0     & 0.24  & 0     & -12   & 1 \\
     2MASS J00085797+0256420        & 2.24154 & 2.945 &  EA                               & 4.72277 & 0.281 & 2012/10/12 &          F0  & 7000  & 20    & 3.96  & 0.01  & -0.06 & 0.01  & -10   & 2 \\
     2MASS J00085797+0256420        & 2.24154 & 2.945 &  EA                               & 4.72277 & 0.08  & 2012/10/29 &          F0  & 7040  & 10    & 3.92  & 0.01  & -0.04 & 0.01  & -10   & 2 \\
     V0468 And                      & 2.44383 & 40.193 &  EA                               & 12.757 & 0.093 & 2013/11/14 &          G6  & 5240  & 10    & 2.57  & 0     & -0.29 & 0     & -76   & 1 \\
     V0468 And                      & 2.44383 & 40.193 &  EA                               & 12.757 & 0.093 & 2013/11/14 &          G7  & 5250  & 10    & 2.65  & 0     & -0.28 & 0.01  & -76   & 2 \\
     2MASS J00121564+3351121        & 3.06517 & 33.85336 &  EA                               & 6.55222 & 0.022 & 2015/9/25 &          F0  & 6620  & 10    & 3.99  & 0     & -0.16 & 0     & -8    & 1 \\
     CSS\_J001241.8+224144           & 3.1745 & 22.69561 &  EA                               & 1.06851 & 0.51  & 2016/12/4 &          G9  & 5306.47 & 90.3  & 4.459 & 0.128 & -0.065 & 0.084 & -23.21 & 7.31 \\
     VSX J001303.0+375505           & 3.26254 & 37.91819 &  EA                               & 0.70804 & 0.151 & 2013/12/15 &          G1  & 5850  & 20    & 4.08  & 0.01  & -0.05 & 0.01  & 8     & 2 \\
     CSS\_J001303.9+030153           & 3.26629 & 3.03142 &  EA                               & 0.398939 & 0.215 & 2013/11/21 &          K3  & 4710  & 110   & 4.2   & 0.14  & 0.15  & 0.1   & -41   & 9 \\
     NSVS 6325565                   & 3.6093 & 36.21364 &  EA                               & 1.8695 & 0.037 & 2015/9/25 &          F5  & 6250  & 30    & 4.21  & 0.03  & 0     & 0.02  & -92   & 3 \\
     NSVS 6325565                   & 3.6093 & 36.21364 &  EA                               & 1.8695 & 0.118 & 2012/11/30 &          F8  & 6110  & 230   & 4.16  & 0.32  & 0.03  & 0.21  & -93   & 17 \\
     NSVS 6325565                   & 3.6093 & 36.21364 &  EA                               & 1.8695 & 0.118 & 2013/12/15 &          F8  & 6220  & 40    & 4.18  & 0.04  & 0.02  & 0.03  & 28    & 3 \\
     2MASS J00144658+3016462        & 3.69408 & 30.2795 &  EA                               & 2.16132 & 0.063 & 2012/11/25 &          F6  & 6290  & 10    & 4.3   & 0     & -0.26 & 0     & 8     & 1 \\
     2MASS J00152309+3257082        & 3.84621 & 32.95228 &  EA                               & 1.79057 & 0.101 & 2012/11/25 &          F4  & 6300  & 160   & 4.22  & 0.23  & -0.71 & 0.15  & -48   & 13 \\
     EY Psc                         & 3.91167 & 18.90125 &  EA                               & 3.12945 & 0.125 & 2015/10/4 &          F5  & 6320  & 10    & 4.18  & 0     & -0.36 & 0     & -67   & 1 \\
     V0544 Cas                      & 4.01171 & 48.91825 &  EA/SD                            & 3.37472 & 0.493 & 2013/10/30 &          F7  & 6310  & 20    & 3.84  & 0.02  & 0.26  & 0.01  & -53   & 2 \\
     CSS\_J001753.5+334826           & 4.47312 & 33.80728 &  EA                               & 0.65454 & 0.017 & 2014/11/22 &          G9  & 5200  & 280   & 4.34  & 0.4   & -0.56 & 0.26  & 19    & 22 \\
     CSS\_J001753.5+334826           & 4.47312 & 33.80728 &  EA                               & 0.65454 & 0.017 & 2014/12/23 &          G8  & 5040  & 200   & 4.26  & 0.28  & -0.67 & 0.18  & -83   & 15 \\
     CSS\_J001804.7+242600           & 4.51983 & 24.43353 &  EA                               & 0.430722 & 0.37  & 2013/10/10 &          G7  & 5240  & 280   & 3.91  & 0.4   & -0.65 & 0.26  & -12   & 21 \\
     CSS\_J001845.4+401614           & 4.68933 & 40.27081 &  EA                               & 0.659822 & 0.247 & 2016/11/1 &          K4  & 4647.43 & 254.91 & 4.257 & 0.366 & -0.297 & 0.237 & 67.89 & 18.67 \\
     CSS\_J001845.4+401614           & 4.68933 & 40.27081 &  EA                               & 0.659822 & 0.017 & 2015/10/17 &          K5  & 4470  & 70    & 4.29  & 0.09  & -0.4  & 0.06  & -54   & 5 \\
     NSVS 3699035                   & 4.88325 & 48.89747 &  EA                               & 0.837858 & 0.059 & 2013/10/30 &          F0  & 6550  & 20    & 4.18  & 0.02  & -0.15 & 0.02  & -43   & 2 \\
     CSS\_J001949.9+232058           & 4.95821 & 23.34969 &  EA                               & 1.161927 & 0.261 & 2016/12/4 &          F3  & 6516.54 & 273.3 & 4.284 & 0.391 & -0.371 & 0.254 & -96.01 & 20.84 \\
     CSS\_J002020.4+331410           & 5.08504 & 33.23636 &  EA                               & 2.51995 & 0.006 & 2014/12/23 &          K0  & 4850  & 130   & 3.65  & 0.17  & -0.45 & 0.11  & -73   & 10 \\
     NSV 131                        & 5.12575 & 44.85303 &  EA:                              &                    & 0.092 & 2016/10/26 &          F0  & 7280.33 & 4.07  & 3.968 & 0.003 & 0.266 & 0.003 & -20.04 & 0.97 \\
     CSS\_J002148.5+205512           & 5.45221 & 20.92017 &  EA                               & 1.13729 & 0.32  & 2015/10/4 &          K0  & 5210  & 130   & 4.24  & 0.17  & 0.05  & 0.11  & 81    & 10 \\
     VSX J002236.2+475641           & 5.65107 & 47.94483 &  EA                               & 0.6322 & 0.078 & 2014/12/20 &          F7  & 6290  & 330   & 4.26  & 0.47  & 0.02  & 0.3   & -23   & 24 \\
     CSS\_J002620.0+270834           & 6.58362 & 27.143 &  EA                               & 0.764235 & 0.855 & 2013/10/10 &        A6IV  & 7170  & 20    & 4.23  & 0.03  & -0.41 & 0.02  & -52   & 3 \\
     CSS\_J002641.1+415921           & 6.67146 & 41.98928 &  EA                               & 0.711088 & 0.574 & 2012/12/4 &          K7  & 4350  & 180   & 4.5   & 0.25  & -0.35 & 0.16  & -14   & 14 \\
     CSS\_J002659.1+350322           & 6.74646 & 35.05625 &  EA                               & 1.48677 & 0.518 & 2016/9/9 &         A7V  & 7513.02 & 6.16  & 4.338 & 0.006 & -0.105 & 0.006 & -18.78 & 1.4 \\
     CSS\_J002659.1+350322           & 6.74646 & 35.05625 &  EA                               & 1.48677 & 0.005 & 2014/11/22 &          F0  & 6850  & 310   & 3.87  & 0.44  & -0.14 & 0.29  & -35   & 22 \\
     ROTSE1 J003005.66+344643.0     & 7.52358 & 34.77861 &  EA                               & 0.948832 & 0.279 & 2012/10/31 &          F2  & 5960  & 110   & 4.23  & 0.15  & -0.44 & 0.1   & -90   & 9 \\
     ROTSE1 J003005.66+344643.0     & 7.52358 & 34.77861 &  EA                               & 0.948832 & 0.279 & 2014/11/22 &          F2  & 6040  & 20    & 4.31  & 0.02  & -0.4  & 0.01  & -19   & 2 \\
     ROTSE1 J003005.66+344643.0     & 7.52358 & 34.77861 &  EA                               & 0.948832 & 0.077 & 2012/11/23 &          F7  & 5870  & 140   & 4.09  & 0.2   & -0.73 & 0.13  & -153  & 11 \\
     NSVS 3777063                   & 7.73754 & 43.16925 &  EA                               & 4.3287 & 0.08  & 2016/10/26 &          F5  & 6270.72 & 48.41 & 4.229 & 0.068 & 0.06  & 0.046 & -13.39 & 4.53 \\
     CSS\_J003258.4+352109           & 8.24338 & 35.35261 &  EA                               & 0.338148 & 0.015 & 2014/10/21 &          G7  & 5120  &           & 3.88  &            & -0.61 &            & -151  &            \\
\hline\hline
\end{tabular}
\end{center}
\end{sidewaystable}

\begin{table*}[h]
\footnotesize
\begin{center}
\caption{Spectral types of EAs determined by LAMOST (the first 50 observations).}\label{XXXX}
\begin{tabular}{llllllll}\hline\hline
Name & R.A. (deg) & Dec. (deg) & Type &  Period (days) & Distance & Date & Sp.\\\hline
    DM Peg                         & 0.03037 & 18.73808 & EA/D:                            & 2.588991 & 0.045 & 2012/9/28 &       A8III   \\
    AM And                         & 1.28137 & 48.4515 & EA                               & 8.85051 & 0.489 & 2013/10/30 &        A3IV   \\
    NSV 15050                      & 3.30354 & 7.96808 & EA                               & 1.665728 & 0.229 & 2013/11/5 &         A5V   \\
    NSV 15050                      & 3.30354 & 7.96808 & EA                               & 1.665728 & 0.229 & 2014/9/18 &        A6IV   \\
    NSVS 6325565                   & 3.6093 & 36.21364 & EA                               & 1.8695 & 0.118 & 2012/11/30 &          F6   \\
    NSVS 6325565                   & 3.6093 & 36.21364 & EA                               & 1.8695 & 0.118 & 2012/11/30 &          F8   \\
    NSVS 6330298                   & 4.76756 & 31.90486 & EA                               & 18.14699 & 1.001 & 2012/11/25 &         A7V   \\
    CSS\_J002620.0+270834           & 6.58362 & 27.143 & EA                               & 0.764235 & 0.743 & 2012/11/24 &        A7IV   \\
    CSS\_J002641.1+415921           & 6.67146 & 41.98928 & EA                               & 0.711088 & 0.599 & 2016/10/26 &          K5   \\
    CSS\_J002659.1+350322           & 6.74646 & 35.05625 & EA                               & 1.48677 & 0.604 & 2012/11/23 &          K3   \\
    CSS\_J002911.3+374322           & 7.29737 & 37.72281 & EA                               & 0.599814 & 0.084 & 2015/10/16 &          M2   \\
    CSS\_J003143.9+384224           & 7.93325 & 38.70689 & EA                               & 4.527844 & 0.525 & 2012/10/31 &          M3   \\
    CSS\_J003143.9+384224           & 7.93325 & 38.70689 & EA                               & 4.527844 & 0.498 & 2014/9/10 &        A1IV   \\
    HR And                         & 8.4445 & 44.06939 & EA                               & 1.2357 & 0.6   & 2012/10/6 &          F5   \\
    FV Cas                         & 9.15171 & 55.22553 & EA/SD                            & 3.06673 & 0.113 & 2013/10/26 &         A5V   \\
    NSVS 1706308                   & 9.68887 & 57.83644 & EA                               & 1.39778 & 0.297 & 2016/9/19 &         A5V   \\
    V0496 And                      & 9.88762 & 27.50814 & EA                               & 4.40262 & 0.245 & 2013/9/15 &         A7V   \\
    V0496 And                      & 9.88762 & 27.50814 & EA                               & 4.40262 & 0.245 & 2013/9/15 &         A7V   \\
    V0496 And                      & 9.88762 & 27.50814 & EA                               & 4.40262 & 0.13  & 2012/10/4 &         A7V   \\
    CSS\_J003949.0+252010           & 9.95454 & 25.33628 & EA                               & 0.870583 & 1.571 & 2012/10/4 &          K7   \\
    V1046 Cas                      & 10.18425 & 58.84828 & EA                               & 0.9805 & 0.098 & 2016/9/19 &        A1IV   \\
    CSS\_J004219.1+390044           & 10.57987 & 39.01236 & EA                               & 0.84472 & 0.014 & 2015/9/13 &          G7   \\
    Mis V1398                      & 11.23729 & 57.07208 & EA:                              &                    & 0.225 & 2013/10/26 &         A9V   \\
    V1049 Cas                      & 11.35417 & 58.09775 & EA                               & 2.98128 & 0.172 & 2016/9/19 &         A1V   \\
    WX And                         & 11.40558 & 28.75 & EA                               & 3.001134 & 0.061 & 2012/9/29 &         A9V   \\
    PTFEB11.441                    & 11.44167 & 41.84167 & EA                               & 0.35871 & 0.384 & 2011/10/28 &          M2   \\
    IL And                         & 11.55633 & 39.76897 & EA                               & 0.86759 & 0.363 & 2011/11/8 &          F3   \\
    IL And                         & 11.55633 & 39.76897 & EA                               & 0.86759 & 0.363 & 2011/11/20 &         A9V   \\
    CSS\_J004726.1+371327           & 11.859 & 37.22428 & EA                               & 1.169599 & 0.196 & 2011/12/12 &          F5   \\
    CSS\_J004726.1+371327           & 11.859 & 37.22428 & EA                               & 1.169599 & 0.196 & 2012/10/3 &          G2   \\
    CSS\_J004828.1+283320           & 12.11708 & 28.55564 & EA                               & 0.867343 & 0.373 & 2013/9/15 &          K5   \\
    CSS\_J005659.0+163753           & 14.24617 & 16.6315 & EA                               & 1.60474 & 1.048 & 2012/10/27 &          K3   \\
    AT Psc                         & 14.61875 & 31.67917 & EA                               & 3.783485 & 0.282 & 2014/11/9 &         A2V   \\
    CSS\_J005906.3+413410           & 14.77654 & 41.56944 & EA                               & 0.971055 & 0.902 & 2012/10/5 &         A5V   \\
    V0386 Cas                      & 14.79696 & 55.95553 & EA/DS                            & 28.65225 & 0.229 & 2016/9/19 &        A7IV   \\
    CSS\_J005922.2+401932           & 14.84258 & 40.32569 & EA                               & 0.796466 & 0.233 & 2015/10/29 &          F9   \\
    CSS\_J010154.3+222307           & 15.47654 & 22.38539 & EA                               & 0.594672 & 1.06  & 2013/10/15 &          K1   \\
    CSS\_J010308.9+381802           & 15.78717 & 38.30064 & EA                               & 0.70891 & 0.312 & 2012/10/7 &          F0   \\
    CSS\_J010308.9+381802           & 15.78717 & 38.30064 & EA                               & 0.70891 & 0.23  & 2016/12/9 &        A6IV   \\
    CSS\_J010308.9+381802           & 15.78717 & 38.30064 & EA                               & 0.70891 & 0.009 & 2015/10/29 &        A6IV   \\
    CSS\_J010308.9+381802           & 15.78717 & 38.30064 & EA                               & 0.70891 & 0.312 & 2015/10/29 &        A6IV   \\
    T-And0-01203                   & 15.89477 & 48.54424 & EA                               & 3.50526 & 0.012 & 2012/11/24 &        A2IV   \\
    BE Psc                         & 16.02979 & 26.58703 & EA/RS                            & 35.67142 & 0.526 & 2012/10/29 &          G9   \\
    V1273 Cas                      & 16.47242 & 53.93522 & EA                               &                    & 0.866 & 2014/11/4 &          F3   \\
    NSVS 6404973                   & 16.75254 & 25.19422 & EA                               & 2.28382 & 1.33  & 2011/12/3 &         A7V   \\
    FK And                         & 16.763 & 37.48481 & EA                               & 2.26941 & 0.766 & 2012/11/23 &        A2IV   \\
    FK And                         & 16.763 & 37.48481 & EA                               & 2.26941 & 0.863 & 2015/12/21 &        A2IV   \\
    CSS\_J010717.9+260500           & 16.82463 & 26.0835 & EA                               & 1.03122 & 0.216 & 2011/12/3 &          K5   \\
    CSS\_J010912.1+291113           & 17.30058 & 29.187 & EA                               & 0.436716 & 1.39  & 2012/10/29 &          G7   \\
    CSS\_J011101.7+313248           & 17.75713 & 31.54686 & EA                               & 1.83522 & 0.68  & 2012/12/25 &          F6   \\
\hline\hline
\end{tabular}
\end{center}
\end{table*}

\section{Distributions of stellar atmospheric parameters for EAs}

\begin{figure}
\begin{center}
\includegraphics[angle=0,scale=0.6]{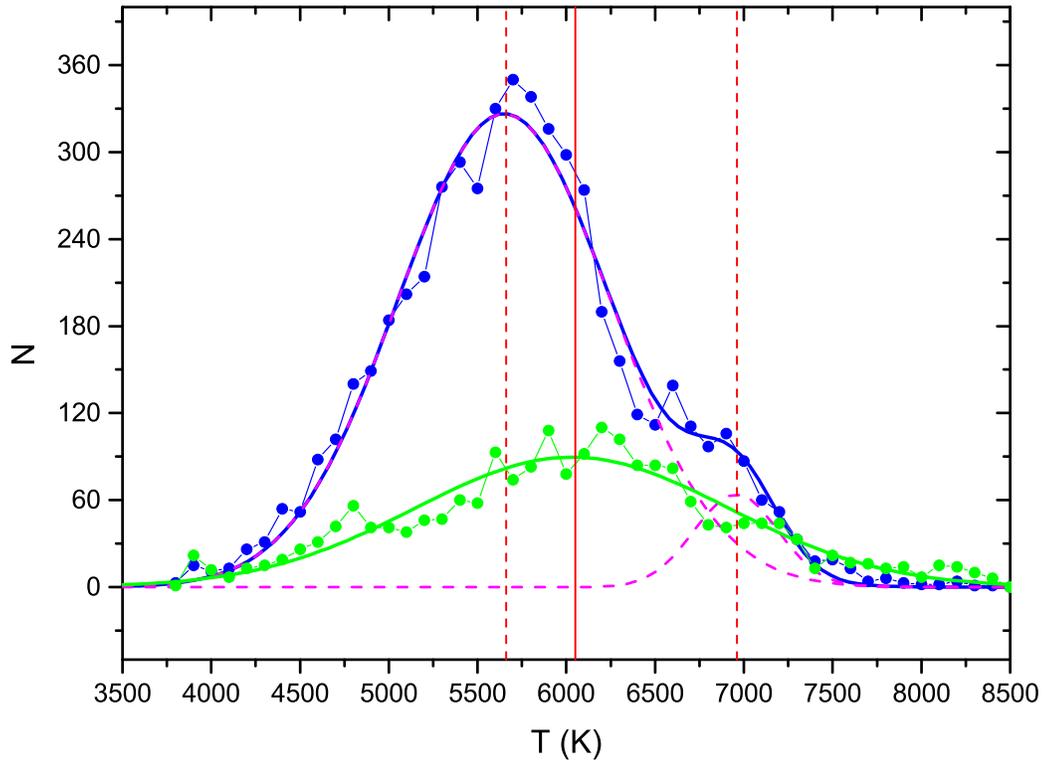}
\caption{Effective temperature distribution for EAs observed by LAMOST (solid green dots). Also shown as blue ones are the distribution of effective temperature for EWs observed by LAMOST. The red solid line refers to the peak for EAs around 6050\,K, while the two dashed lines represent the peaks for EWa near 5660 and 6960\,K, respectively.}
\end{center}
\end{figure}

\begin{figure}
\begin{center}
\includegraphics[angle=0,scale=0.6]{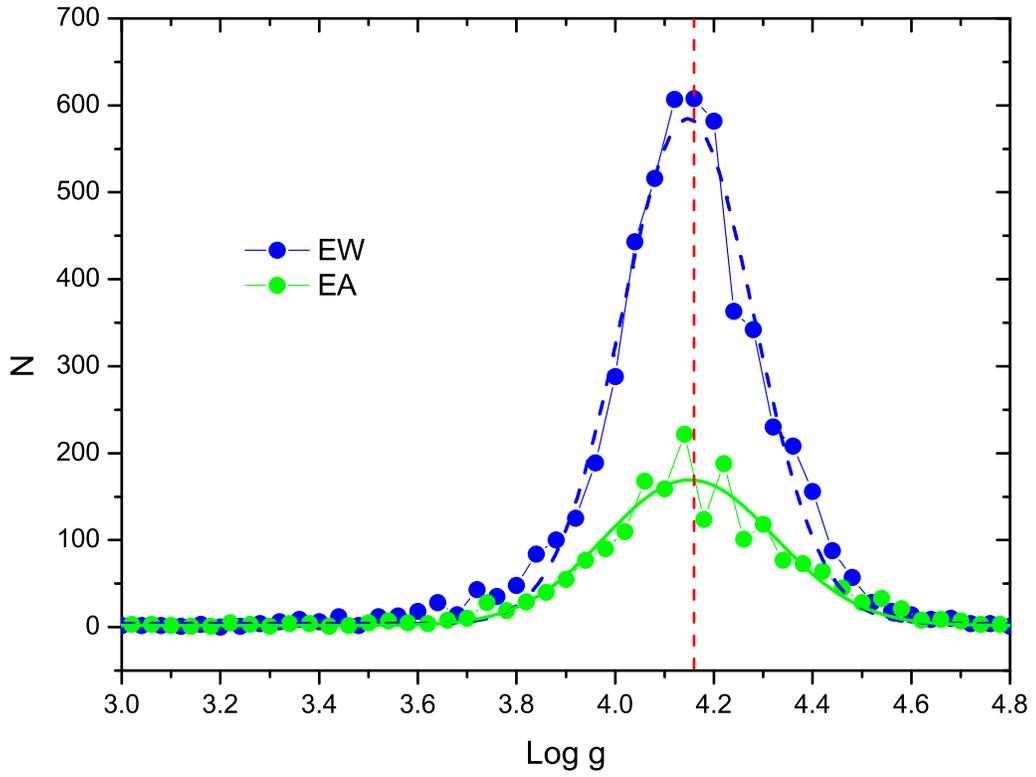}
\caption{Distributions of the gravitational acceleration Log (g) for EAs and EWs observed by LAMOST. The red dashed line refers to the peak around 4.16.}
\end{center}
\end{figure}

\begin{figure}
\begin{center}
\includegraphics[angle=0,scale=0.6]{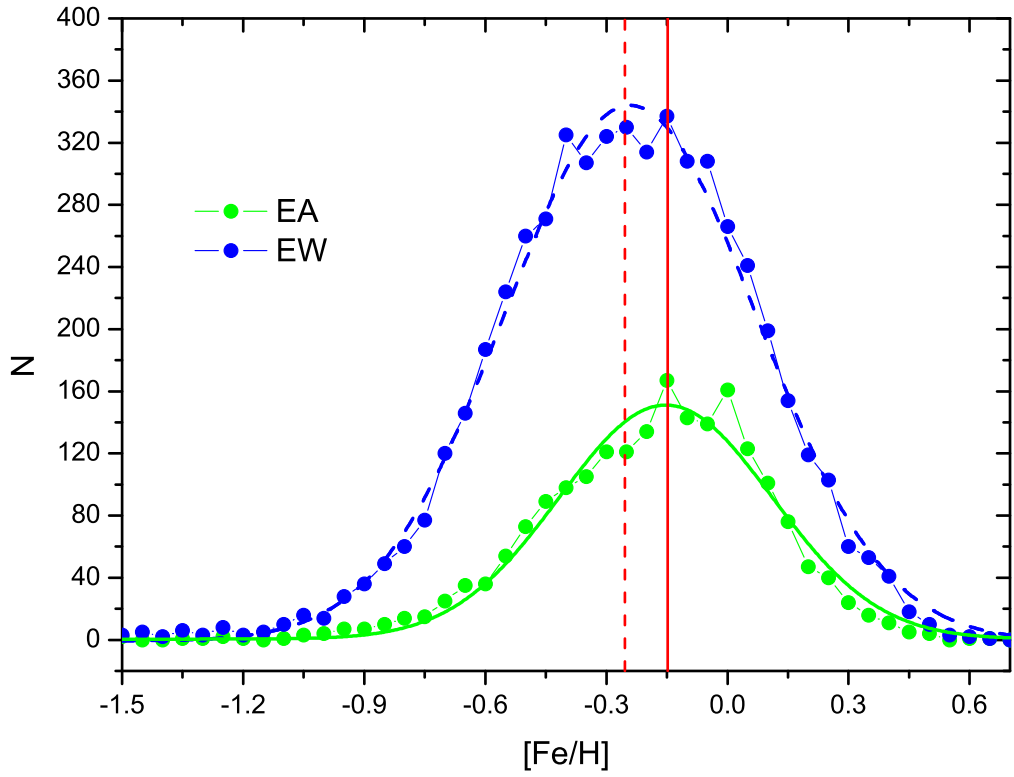}
\caption{Distributions of the metallicity [Fe/H] for EAs and EWs observed by LAMOST. The peak for EAs is near -0.15, while that for EWs is near -0.24.}
\end{center}
\end{figure}

\begin{figure}
\begin{center}
\includegraphics[angle=0,scale=0.6]{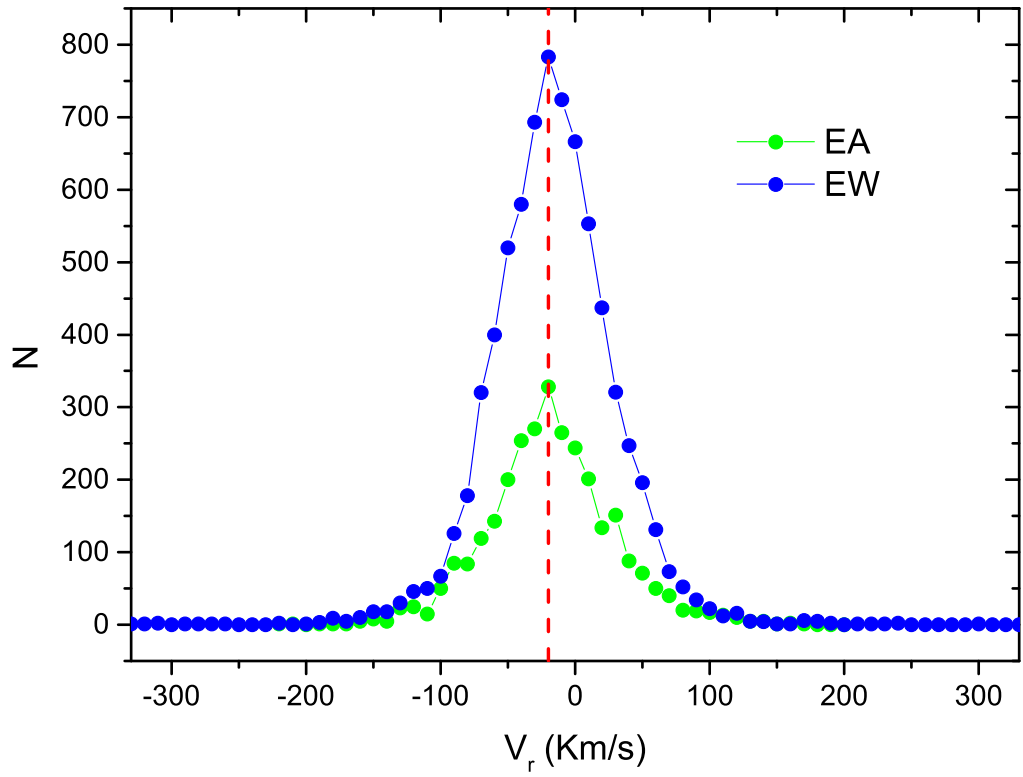}
\caption{Distributions of the radial velocity $V_{r}$ for EAs and EWs observed by LAMOST. The peak is near $V_{r}=-20$\,Km/s for both EAs and EWs.}
\end{center}
\end{figure}

To analyze the distributions of the effective temperature $T$, the gravitational acceleration Log g and the metallicity [Fe/H], when the EA samples were observed two times or more, we average those stellar atmospheric parameters and use the mean values. For each observation, the weight is the inverse square of the original error. As for the radial velocity $V_{r}$, we do not average them and use the individuals because they are varying with time. The temperature distribution is shown in Fig. 4. For comparison, the temperature distribution of EWs is also displayed in the same figure. The distribution peak for EAs is near 6050\,K, while there are two peaks near 5660\,K and 6960\,K for EWs that have been pointed out by Qian et al. (2017a). The main peak of EWs reveals that most cool EWs have temperatures lower than those of EAs indicating that they may have a long lifetime in contact stage. The second small peak of EWs shows that there is a group of hotter EWs whose temperatures are higher than most those of EAs. They may be formed from EAs via mass transfer.

The distribution of the gravitational acceleration Log g for EAs is plotted in Fig. 5 as solid dots. Also displayed in the same figure as open circles are the distribution of EWs. It is shown that the peaks of the two distributions are near 4.16. This reveals that both the samples of EAs and EWs are main-sequence binaries according to the fundamental parameters listed in Table VII of the paper published by Straizys \& Kuriliene (1981). If EAs evolve into EWs through the mass transfer from the primary to the secondary, the mass transfer should be occurred during the main-sequence stage. On the other hand, the distributions may suggest that EWs may be formed from original cool main-sequence short-period EAs via angular momentum loss. These are consistent with the assumption that EWs are evolved from EAs through angular momentum loss via magnetic braking and/or Case A mass transfer (e.g., Bradstreet \& Guinan 1994; Qian et al. 2013a, 2017).

The EA metallicity ([Fe/H]) distribution is shown in Fig. 6. The EW distribution is also shown in the same figure. As we can see that the peak for EA-type binaries is near -0.15, while the [Fe/H] peak for EW is near -0.24. Since stellar metallicities are
weakly correlated with their ages for stars in the Galaxy (e.g., Reid et al. 2007; Feltzing \& Bensby 2009), the lower metallicities of {\bf some EWs} may reveal that they are usually older than EAs. The stellar atmospheric parameters of ten EAs with the highest metallicities are shown in Table 4. {\bf However, the mean and the peaks of the EW and EA metallicity distributions look to be consistent within the error of the mean. To check if this is right or not, three independent statistic tests including Kolmogorov-Smirnov test, the Kruskal-Wallis H-test, and the nonparametric Rank-Sum test are performed to analyze the [Fe/H] distributions of EA and EW binaries. Their probabilities are calculated to be $7.7\times{10^{-21}}$, 0, and 0, respectively. They are all much smaller than the 0.05 significance level. These results indicate that the [Fe/H] distributions for EAs and EWs are not statistically consistent.} The distributions of the radial velocity ($V_{r}$) for both EAs and EWs are displayed in Fig. 7. It is shown that both of the two peaks are near $V_{r}=-20$\,Km/s that may reveal that the values of $V_0$ for most EAs and EWs are close to this value. The distributions are symmetric that reflects a statistical random sampling of radial velocity curves for EAs and EWs. Twenty EAs with the largest radial velocities are shown in Table 5. These EAs may be observed by LAMOST near the maxima of the radial velocity curves.

\begin{table}[h]
\footnotesize
\begin{center}
\caption{Ten EAs with the highest metallicities.}\label{XXXX}
\begin{tabular}{llllllll}\hline\hline
Name &  P (days) & Dates & Sp. & T (K) &  Log(g)  & [Fe/H] &  $V_r$ \\\hline
     V2365 Oph                      & 4.86562 & 2017/4/29 &         A5V  & 8272.44 & 4.0     & 0.664 & 65.37 \\
     KID 04737302                   & 9.455584 & 2015/10/12 &          F0  & 7330  & 3.9   & 0.63  & 11 \\
     KID 04737302                   & 9.455584 & 2017/6/13 &          F0  & 7348.52 & 3.883 & 0.624 & -49.63 \\
     KID 04737302                   & 9.455584 & 2012/6/15 &          F0  & 7260  & 3.88  & 0.59  & -86 \\
     CSS\_J091340.2-010907           & 2.021 & 2013/12/20 &          K1  & 5560  & 4.2   & 0.53  & -18 \\
     T-UMa0-00838                   & 1.009849 & 2013/1/28 &          K3  & 4800  & 4.46  & 0.52  & -17 \\
     TYC 1862-2014-1                & 2.0045 & 2013/12/20 &          F0  & 7290  & 3.79  & 0.52  & 28 \\
     CSS\_J091340.2-010907           & 2.021 & 2013/12/20 &          K1  & 5530  & 4.14  & 0.52  & -21 \\
     KID 06449358                   & 5.776786 & 2015/10/18 &          F0  & 7440  & 3.93  & 0.51  & -65 \\
     CSS\_J021626.6+273742           & 1.6679788 & 2013/10/31 &          F3  & 6850  & 3.95  & 0.51  & -11 \\
     KID 05653126                   & 38.494555 & 2014/5/22 &          F0  & 7100  & 4.16  & 0.5   & -39 \\
     KID 02860594                   & 5.499937 & 2012/6/15 &          F0  & 7220  & 3.73  & 0.49  & -66 \\
\hline\hline
\end{tabular}
\end{center}
\end{table}

\begin{table}[h]
\footnotesize
\begin{center}
\caption{Twenty EAs with the largest radial velocities.}\label{XXXX}
\begin{tabular}{lllllll}\hline\hline
Name &  P (days) & Sp. & T (K)& Log(g)
&[Fe/H] &  RV \\\hline
     NSVS 2517147                   & 0.446771 &          K1  & 5170  & 4.07  & -0.13 & -214 \\
     CSS\_J075153.6+550929           & 0.456618 &          K0  & 4790  & 3.85  & -0.69 & -204 \\
     1SWASP J170101.25+490659.3     & 2.807525 &          G3  & 5360  & 3.73  & -0.75 & -189 \\
     CSS\_J042218.1+180600           & 0.766028 &          K7  & 4060  & 4.14  & -0.79 & 176 \\
     CSS\_J162503.4+311034           & 0.793984 &        A6IV  & 6780  & 4.23  & -0.77 & -175 \\
     KID 07943602                   & 14.69202 &          G9  & 4872.6 & 2.866 & -0.717 & -168.85 \\
     CSS\_J093524.1+110835           & 0.55175 &          K1  & 5180  & 4.18  & 0.07  & 163 \\
     FASTT 79                       & 0.709397 &          G2  & 5900  & 4.16  & -0.22 & 160 \\
     CSS\_J160450.1+520159           & 328.5 &          G7  & 5100  & 2.88  & -1.24 & -158 \\
     CSS\_J173216.3+365552           & 0.827061 &          F0  & 6170  & 4.29  & -1.09 & -158 \\
     CSS\_J011542.9+285000           & 1.451186 &          G9  & 5000  & 4.15  & -0.76 & -156 \\
     ROTSE1 J003005.66+344643.0     & 0.948832 &          F7  & 5870  & 4.09  & -0.73 & -153 \\
     CSS\_J003258.4+352109           & 0.338148 &          G7  & 5120  & 3.88  & -0.61 & -151 \\
     CSS\_J091449.2+020342           & 0.739141 &          K3  & 4760  & 4.32  & 0.03  & 150 \\
     V1361 Her                      & 1.34119 &          K4  & 4740  & 4.43  & -0.55 & -149 \\
     CSS\_J013541.9+383916           & 1.27451 &          K3  & 4920  & 4.32  & -0.13 & -149 \\
     CSS\_J220825.5+221514           & 4.541347 &         A6V  & 6720  & 4.33  & -0.86 & -149 \\
     KID 07943602                   & 14.69202 &          K0  & 4780  & 3.31  & -0.68 & -148 \\
     CSS\_J015424.6+390419           & 0.562459 &          K3  & 4850  & 4.45  & -0.56 & -145 \\
     CSS\_J082619.4+192926           & 0.293376 &          K4  & 4590  & 4.16  & -0.72 & 145 \\
\hline\hline
\end{tabular}
\end{center}
\end{table}

\section{Analysis of Binary Property Correlations}

\begin{figure}
\begin{center}
\includegraphics[angle=0,scale=1.2]{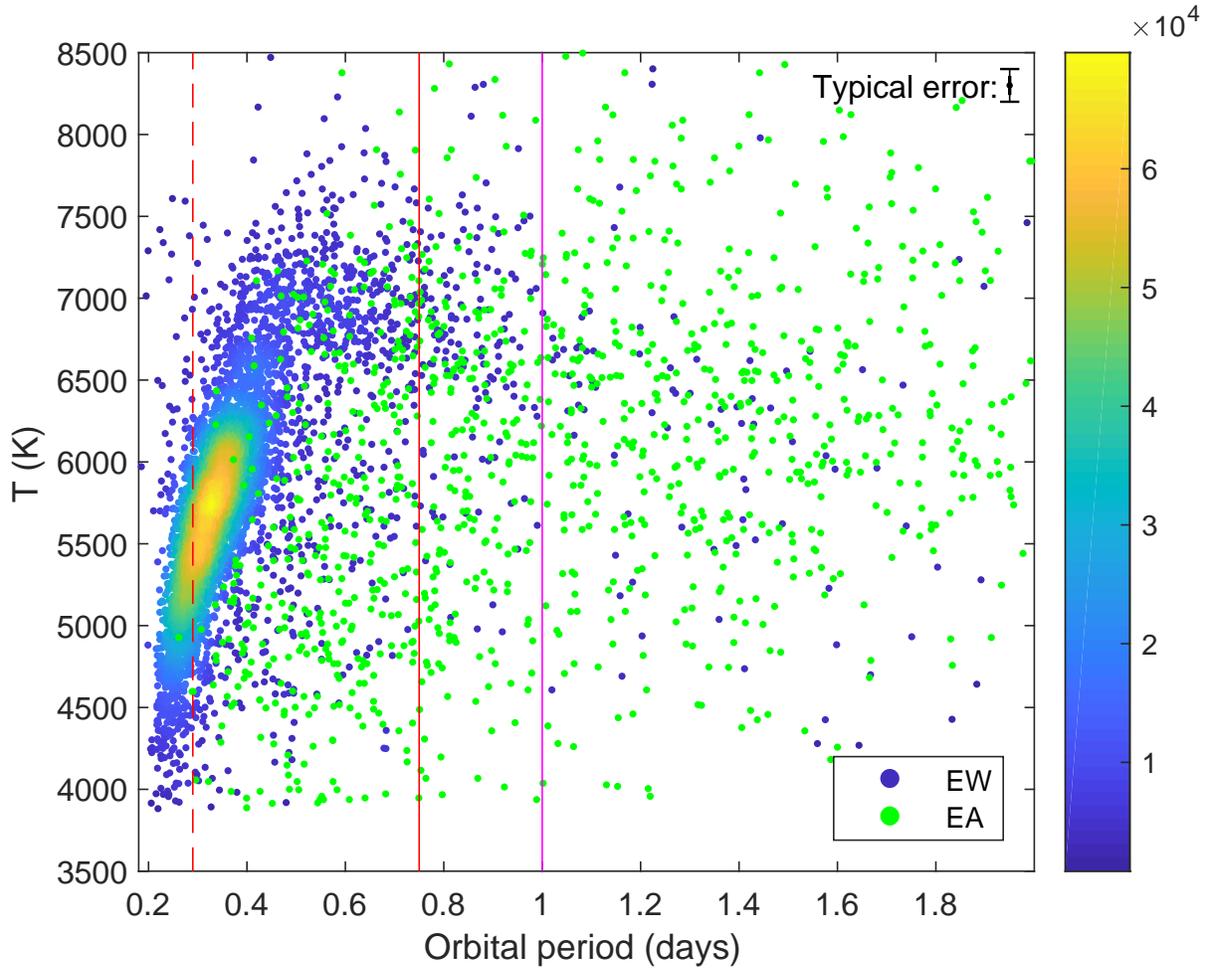}
\caption{The comparison between the orbital period and the effective temperature for EAs and EWs observed by LAMOST. Blue dots refer to EWs, while green ones to EAs. Only binary systems with orbital period shorter than 2 days are shown. The red solid line refers to the period peak for EAs, while the red dashed one to that of EWs. {\bf The magenta line denotes periods of 1 day.}}
\end{center}
\end{figure}

\begin{figure}
\begin{center}
\includegraphics[angle=0,scale=1.2]{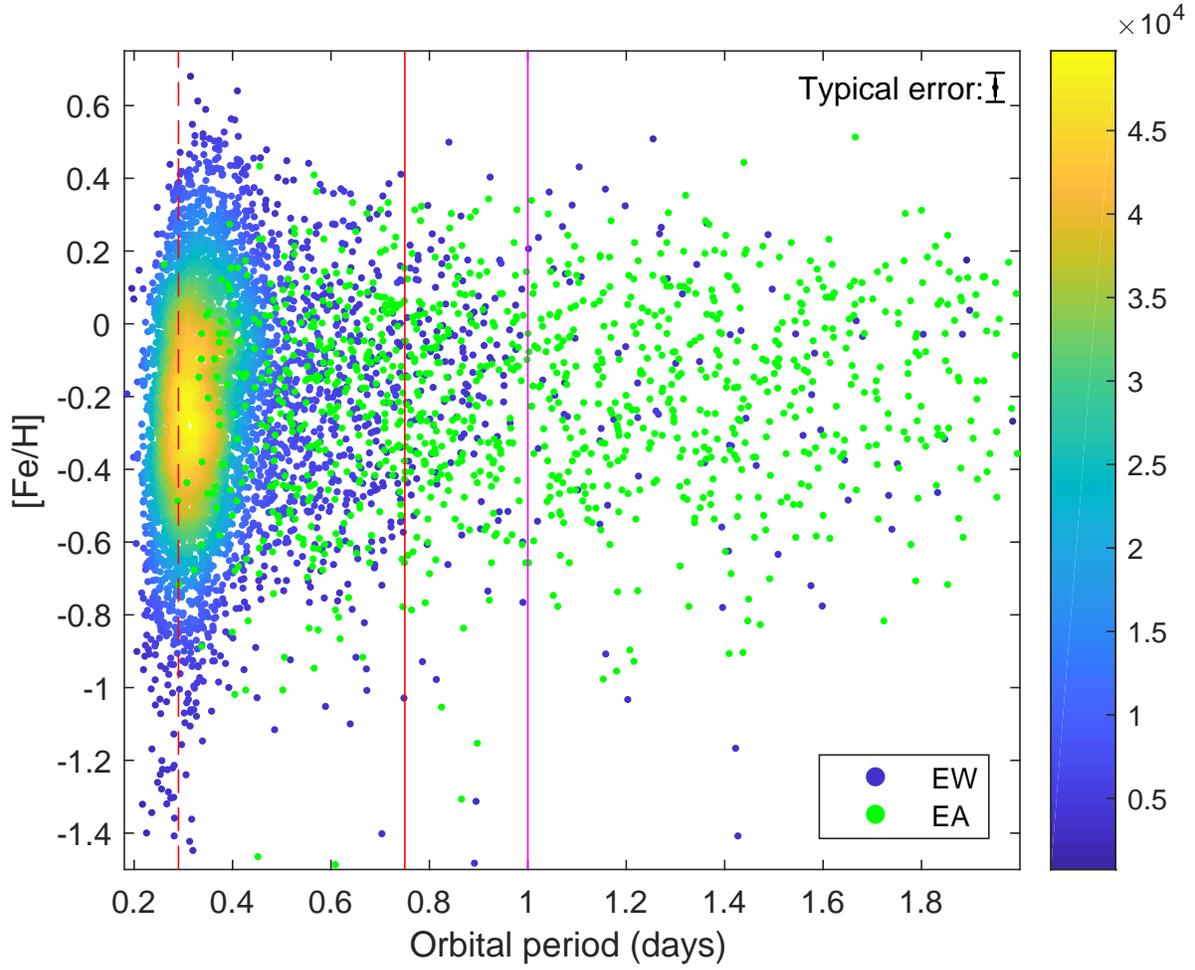}
\caption{The same as those in Fig. 8 but for the comparison between the orbital period and the metallicity.}
\end{center}
\end{figure}

\begin{figure}
\begin{center}
\includegraphics[angle=0,scale=1.2]{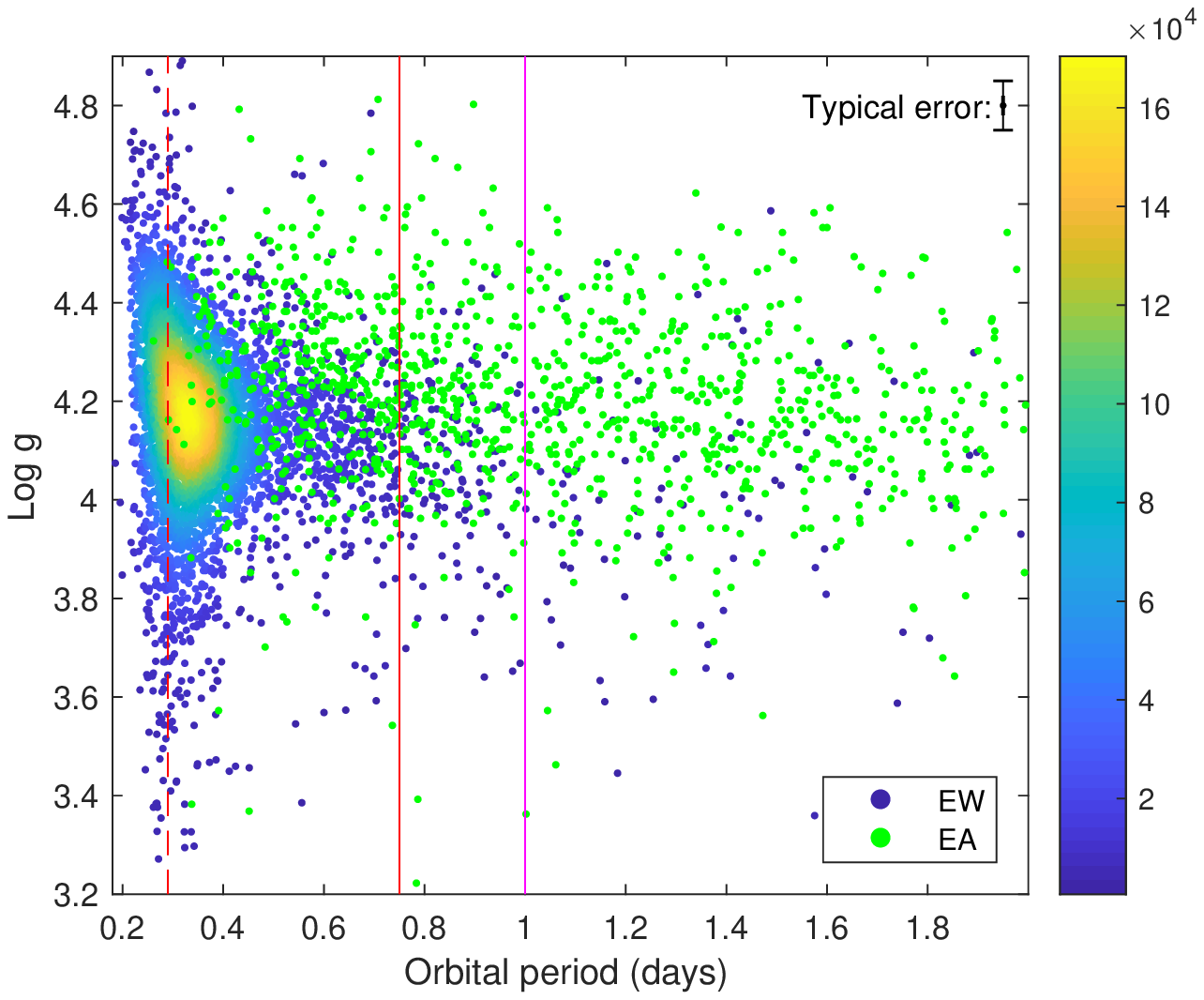}
\caption{
The same as those in Figs. 8 and 9 but for the comparison between the orbital period and the gravitational acceleration.}
\end{center}
\end{figure}

The relation between the orbital period and the effective temperature $T$ for EAs with orbital period shorter than 2\,days is shown in Fig. 8. For comparison, EWs are also displayed in the figure as blue dots. The red solid line refers to the peak of the period distribution for EAs, while the red dashed one to the period peak for EWs. As displayed in the figure, there is a tight correlation between the orbital period and the effective temperature for short-period EWs (e.g., Qian et al. 2017a). However, the effective temperature of EAs is not correlated with the orbital period and the distribution of the samples is uniformed. These are expected results because EWs are usually contact systems where both components are sharing a common envelope, while the components in EAs are not in contact with each other. For hotter EWs with temperature larger 6500\,K, their orbital periods extend to 1 day (the solid magenta line) or longer. A small number of EWs have longer periods when compared with their temperatures indicating that the component stars have been evolved from main sequence and may be subgiants or giants.

Fig. 9 displays the relation between the orbital period and the metallicity [Fe/H] where green dots refer to EAs and blue ones to EWs. It is shown that the positions of some long-period EWs are overlapping with those of EAs. The metallicities of some short-period EWs are lower than those of EAs indicating that they are old targets. These EWs may have a long-term pre-contact evolution. For some EWs and EAs, they are metal rich. Their high metallicities may be caused by the contamination of material from the evolution of an unseen degenerate object, i.e., neutron stars or black holes (e.g., Qian et al. 2008, 2017a). Previous work has shown that EWs usually contain additional bodies (e.g., Pribulla et al. 2006). The other possibility is that they may be young systems. The presence of third bodies may shorten their pre-contact evolution and may help the formation of those young EWs (e.g., Qian et al. 2006, 2007, 2013b; Zhu et al. 2013a, b).

The relations between the orbital period and Log g for EAs and EWs with orbital periods shorter than 2\,days is shown in Fig. 10. As those plotted in Figs. 8 and 9, the dashed lines represent the peak values of the period distributions for EAs and EWs. For EWs, the gravitational acceleration is weakly correlated with the orbital period, while no such correlations exist for EAs. As we can see in Fig. 10, for some longer-period EWs (e.g., $P > 0.4$\,days), their values of Log g are usually smaller than those of the EAs with the same periods. This indicates that the long-period EWs are more evolved than those EAs who are progenitors of those EWs. The evolutions of both components in EAs will cause their expanding and after the primary is filling the critical Roche lobe the material will transfer from the primary to the secondary. At the same time, the orbital period is decreasing if mass and angular momentum ar
e conservative and the Roche lobes will be shrinking. When both components are filling the critical Roche lobes and sharing a common envelope, an EW-type contact system is formed (e.g., Qian 2002, 2003). However, for some short-period EWs, their values of Log g are usually larger than those of the EAs indicating that their component stars have not evolved nearly. These systems may be formed from cool short-period EAs with little mass transfer. Angular momentum loss through magnetic braking may play a more important role for their formations (e.g., Qian et al. 2013a, 2014).

\section{Discussions and conclusions}

\begin{figure}
\begin{center}
\includegraphics[angle=0,scale=0.6]{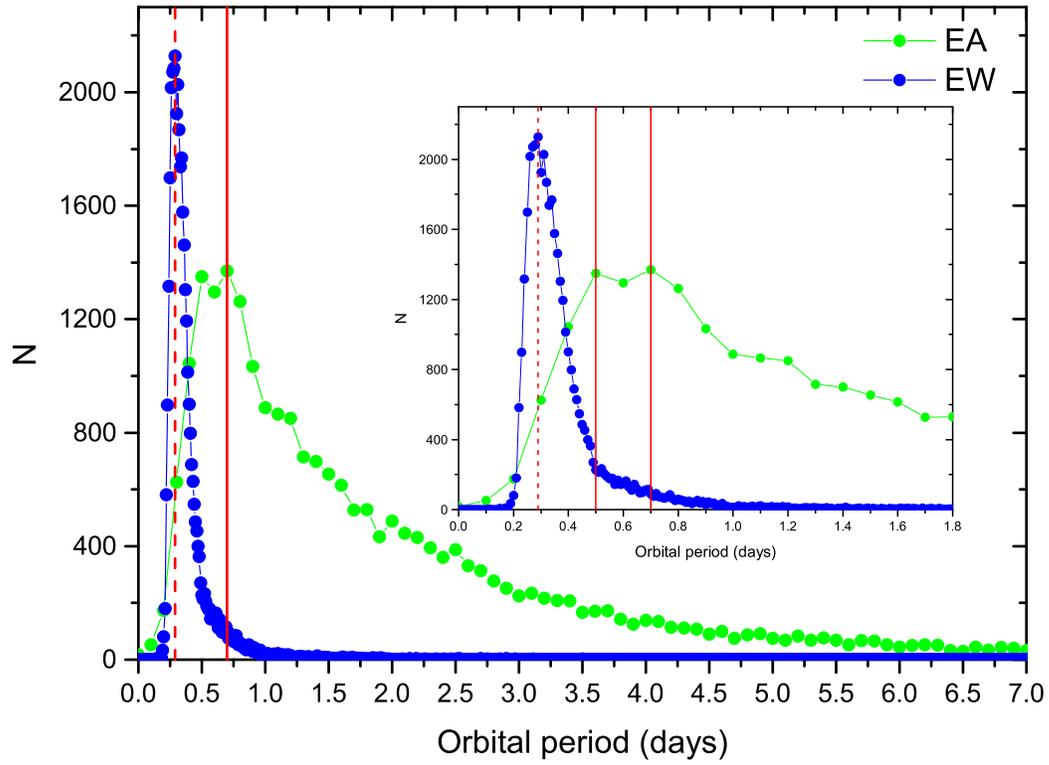}
\caption{Comparison of the period distributions for EAs and EWs. The peak for EA period distribution is around 0.7 days, while that for EW period distribution is near 0.29 days.}
\end{center}
\end{figure}

\begin{figure}
\begin{center}
\includegraphics[angle=0,scale=1.2]{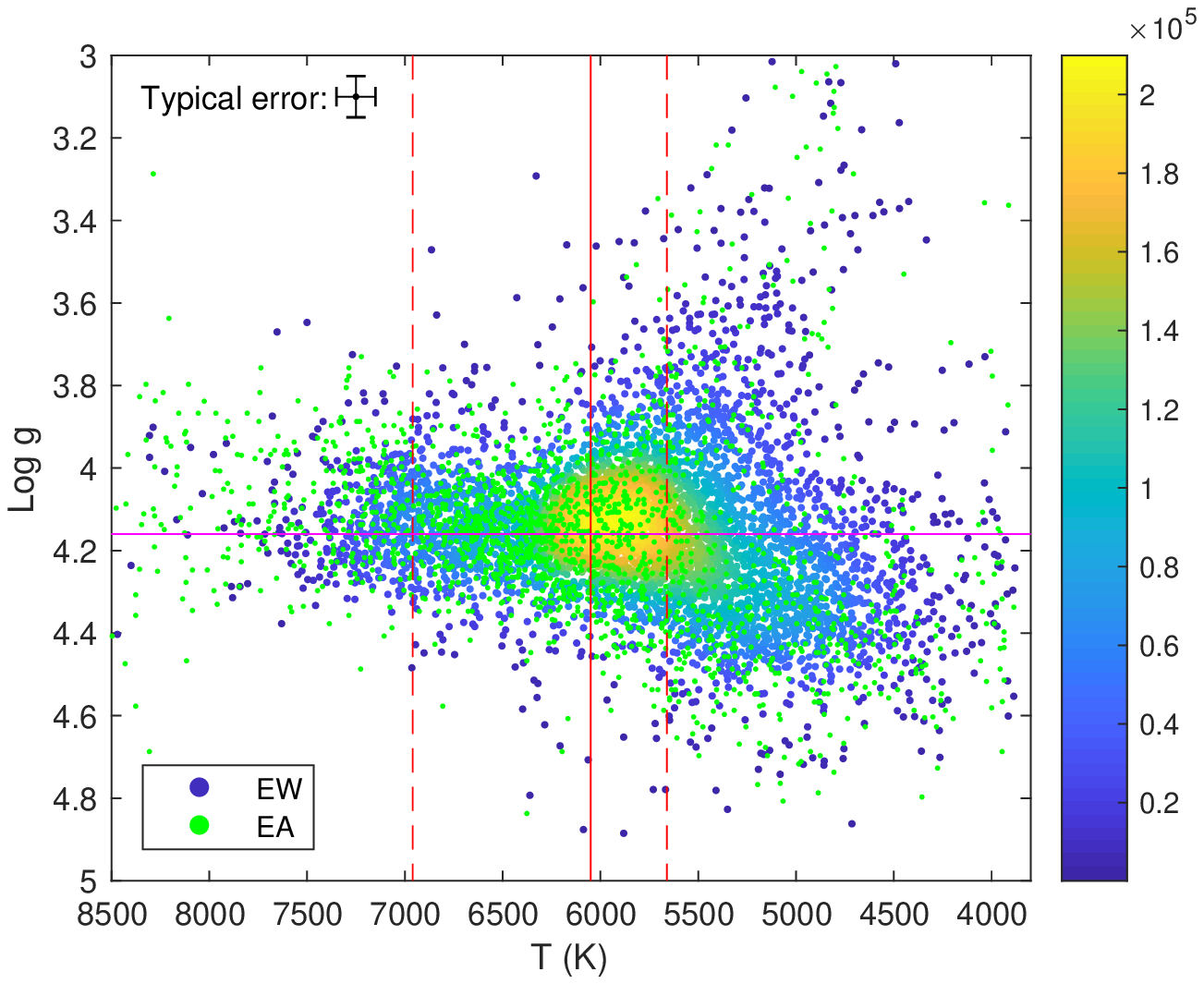}
\caption{The relations between the effective temperature and the gravitational acceleration for EAs and EWs. The red solid and dashed lines refer to the peaks of the temperature distributions for EAs and EWs, respectively. The solid magenta line represents the distribution peak of the gravitational acceleration for both EAs and EWs. Symbols are the same as those in Figs. 8-10.}
\end{center}
\end{figure}

EA-type eclipsing binaries are an important source to determine stellar parameter such as component masses, radii and temperatures etc. They provide a good chance to investigate several astrophysical processes (e.g. binary interacting, mass transfer and magnetic braking etc.) and most importantly they are also the progenitors of EW-type contact binary stars. A huge number of EAs were discovered by several large photometric surveys in the world (e.g., CSS, the asteroid survey LINEAR, ASAS and NSVS). In VSX catalogue, 25364 EAs are listed by July 16, 2017 and about 12.6\% of them (about 3196 EAs) were surveyed by LAMOST from October 24, 2011 to June 16, 2017. Those observed EAs are catalogued and their spectral types are shown. For about 2020 EAs, their stellar atmospheric parameters are also presented. Those effective temperatures as well as the spectral types could be used during solving their light curves. The stellar atmospheric parameters provide us valuable information on their properties and evolutionary states.

{\bf
The stellar atmospheric parameters of 352 stars were determined based on high-resolution optical spectra by Frasca et al. (2016) who collected those high-resolution data from the literature.
The recently released LAMOST data (DR4 and the first three quarters of DR5) have been compared by Qian et al. (2017b) with those derived by Frasca et al. (2016). It is shown that the standard deviations for T and Log g were determined as 135\,K and 0.21 dex, while those for [Fe/H] and $V_{r}$ were derived as 0.14 dex and 11\,Km/s for $V_{r}$, respectively.
The LAMOST binary parameters were investigated by Liu et al. (2017) who measured the combined binary spectra with the same pipeline. They showed that the temperature deviations between the primary stars and the single measured values are mostly less than 200 K, while the deviations for Log g and [Fe/H] are mostly smaller than 0.2 dex. These deviations are usually decreasing with large temperature differences, small mass ratios, or low metallicities. The present EAs and EWs are usually semi-detached and contact binary systems. To check their LAMOST data, the spectral types of 72 EAs and EWs are compared with those determined by previous investigators. As shown in Fig. 2, apart from one EW, all of the other samples are in agreement with previously determined ones within 5 subclasses. Most of them (61 samples) agree within 3 subclasses. In section 2, by analyzing the data of 62 EAs observed four times or more, the standard errors for their parameters were obtained. It is shown that the standard errors for most targets are lower than 100\,K for T, 0.1\,dex for log g and [Fe/H]. As for EWs, similar analyses were done by Qian et al. (2017a). The standard errors are usually lower than 110 K for T, 0.19 dex for log g, and 0.11 dex for [Fe/H], respectively.}

The period distributions for both EAs and EWs shown in VSX are plotted in Fig. 11. It is displayed that the peak of the period distribution for EAs is around 0.7\,days, while the peak for EWs is near 0.29\,days. Both of the two peaks are shorter than those given by previous investigators (e.g., Paczy\'{n}ski 2006) based on the data from ASAS survey. As pointed out by Qian et al. (2017a), the difference may be caused by the reason that ASAS is dedicated to the detection of the variability of bright stars, while many faint eclipsing binaries were discovered by recent deep photometric surveys (e.g., Drake et al. 2009, 2014). As shown in Fig. 11, when the orbital period is shorter than 6\,days, the number of EAs is increasing rapidly. This may indicate that the period limit to tidal locking for the short-period EAs is about 6\,days. In those tidally locked EAs with later type components, the spin angular momentum loss from the components is provided by the reservoir of the orbital angular momentum through the spin-orbit coupling. This causes the orbit to shrink and the orbital period to decrease. In this way, the gradual accumulation of EAs results in the number to increase until the orbital period reach the peak 0.7\,days. The number of EAs is decreasing when the orbital period is shorter than 0.5\,days. As for EWs, when the orbital period is shorter than one day, the number is increasing gradually. Then the number of EWs is increasing rapidly after the orbital period is lower than 0.5\,days. These could be explained as that EAs are evolving into EW through mass transfer during this stage.

The distributions of the effective temperature, the gravitational acceleration, the metallicity and the radial velocity for those EAs are presented. For comparison, the distributions for EWs are also displayed in those figures. The most interesting result is that the peak of the metallicity distribution for EA-type binaries is near -0.15, while the [Fe/H] peak for EW is near -0.24. The lower metallicities for EWs may suggest that they are usually older than EAs. This supports the assumption that the EAs will evolve into EWs through Case A mass transfer and/or AML via magnetic braking. It takes a long-term pre-contact evolution with timescales from a few hundred million to a few billion years.
The relations between the effective temperature and the gravitational acceleration for EAs and EWs are shown in Fig. 12. As we can see in the figure that most EAs and EWs are main-sequence stars and they overlap in the same region. These support the idea that EWs are formed through Case A mass transfer and/or AML. It is interesting to point out that the peaks of radial velocity distributions for both EAs and EWs are near $V_{r}=-20$\,Km/s. This may indicate that the $V_0$ for most EAs and EWs are close to $-20$\,Km/s.

In the previous sections, we analyze the correlations between the orbital period and the effective temperature, the metallicity, and the gravitational acceleration for EAs and EWs. As displayed in Fig. 9, the metallicities of some long-period EWs ($0.4 < P < 1$ days) are the same as those of EAs with the same orbital period. Fig. 10 shows that the values of Log g for some long-period EWs are usually smaller than those of the EAs with the same periods. All of the observational facts support the evolutionary process that the short-period EAs ($P < 6$\,days) will evolve into EWs through the combination of angular momentum loss via magnetic braking and the case A mass transfer.

As shown in Figs. 9 and 10, for some short-period EWs, their metallicities are lower than those of EAs, while their gravitational acceleration are higher. They may be old population systems and their component stars have not evolved nearly. These systems may be formed from short-period EAs with late components through angular momentum loss via magnetic braking with little mass transfer. This kind of binary systems have a long timescale of pre-contact evolution. As displayed in Fig. 9, some EWs have higher metallicities than those of EAs. They may be contaminated by the material from the evolution of unseen neutron stars and black holes in the systems. On the other hand, they may be really very young and have third bodies. Those third component stars help them to form contact configurations and shorten the timescale of pre-contact evolution.

By analyzing the kinematics of 129 field EWs, Bilir et al. (2005) divided EWs into several groups based on their orbital period. The shorter-period less-massive EWs are older than the longer-period more massive systems. Our results are in agreement with those conclusions. As displayed in Figs. 9 and 10, EWs could be divided into at least three groups. The first group are the long-period more massive EWs with the same metallicities but lower Log g as those of EAs with the same orbital period. They are formed from EAs mainly via Case-A mass transfer. The second group are short-period less-massive EWs that have lower metallicities and higher Log g. They are formed from short-period cool EAs mainly through AML via magnetic braking. The last group of EWs have higher metallicities that may be contaminated by the material from the evolution of compact objects or third bodies may play a main role for their formation. The present investigation indicates that the evolution of EAs and the formation of EWs are more complex than we thought. Those EWs are formed through the combination of several mechanisms.

\acknowledgments{This work is partly supported by Chinese Natural Science Foundation (No. 11325315). Guoshoujing Telescope (the Large Sky Area Multi-Object Fiber Spectroscopic Telescope LAMOST) is a National Major Scientific Project built by the Chinese Academy of Sciences. Funding for the project has been provided by the National Development and Reform Commission. LAMOST is operated and managed by the National Astronomical Observatories, Chinese Academy of Sciences. Spectroscopic observations of those EAs used in the paper were obtained with LAMOST from October 24, 2011 to June 16, 2017.}

\end{document}